\documentclass{article}

\usepackage{arxiv}

\usepackage[utf8]{inputenc} 
\usepackage[T1]{fontenc}    
\usepackage{hyperref}       
\usepackage{url}            
\usepackage{booktabs}       
\usepackage{amsfonts}       
\usepackage{nicefrac}       
\usepackage{microtype}      
\usepackage{lipsum}
\usepackage{amssymb}
\usepackage{amsmath}
\usepackage{bm}
\usepackage{stmaryrd}
\usepackage{placeins}
\usepackage[lofdepth,lotdepth]{subfig}
\usepackage{graphicx}
\graphicspath{{./figs/}}
\usepackage[titletoc]{appendix}
\title{Vector Quantized Contrastive Predictive Coding for Template-based Music Generation}

\author{
  Ga\"etan Hadjeres\thanks{Equal contribution} \\
  Sony Computer Science Laboratories\\
  Paris, France\\
  \texttt{gaetan.hadjeres@sony.com} \\
   \And
  Léopold Crestel${}^*$\\
  Sony Computer Science Laboratories\\
  Paris, France\\
  \texttt{leopold.crestel@sony.com} \\
}
\date{}
\begin{document}
\maketitle
\begin{abstract}
In this work, we propose a flexible method for generating variations of discrete sequences in which tokens can be grouped into basic units, like sentences in a text or bars in music.
More precisely, given a template sequence, we aim at producing novel sequences sharing perceptible similarities with the original template without relying on any annotation; so our problem of generating variations is intimately linked to the problem of learning relevant high-level representations without supervision.
Our contribution is two-fold:
First, we propose a self-supervised encoding technique, named \emph{Vector Quantized Contrastive Predictive Coding}
which allows to learn a meaningful assignment of the basic units over a discrete set of codes, together with  mechanisms allowing to control the information content of these learnt discrete representations.
Secondly, we show how these compressed representations can be used to generate variations of a template sequence by using an appropriate attention pattern in the Transformer architecture. 
We illustrate our approach on the corpus of J.S. Bach chorales where we discuss the musical meaning of the learnt discrete codes and show that our proposed method allows to generate coherent and high-quality variations of a given template.
\end{abstract}


\section{Introduction}
\label{sec:introduction}











Generative models for sequences of discrete tokens have been shown to be central in many recent applications: with enough capacity, they are able to model the data distribution of a wide range of data types such as text \cite{vaswani2017attention,Child2019}, images (treated as a single sequence of pixels) \cite{van2017vqvae,salimans2017pixelcnn++,chen2017pixelsnail} or symbolic music \cite{huang2018music,choi2019encoding,Payne2019,donahue2019piano,donahue2019lakhnes}.
Most of the recent progress in this domain builds upon the Transformer architecture from \cite{vaswani2017attention} and especially on the multi-head self-attention introduced therein. Further improvements were allowed by the addition of an efficient relative self-attention mechanism \cite{shaw2018self,huang2018music}, the replacement of the point-wise feedforward neural network with a memory layer \cite{lample2019}, the introduction of sparse attention \cite{correia2019adaptively,sukhbaatar2019adaptive,Child2019,kitaev2020reformer} or by scaling up the model \cite{Child2019,dai2019transformer,kitaev2020reformer}.

Even though these models can generate high quality material, we believe that the limited control they provide could hamper their use for creative applications.
Indeed, when using these autoregressive models in an interactive setting, the possibilities offered to a user to control the generation are often restricted to specifying a primer sequence \cite{chen2017pixelsnail} or imposing a predefined class \cite{razavi2019generating}. 


In this paper, we draw our attention on the problem of generating variations based on a given template sequence as this would provide an additional path for a user to generate new material in an intended way. 
A type of data where this problem is of particular interest is symbolic music, which can be seen as sequences of discrete note events \cite{briot2019deep}. Being able to compose variations of a given piece is a natural musical practice (e.g. theme and variations) and useful for creative usages.


The problem of generating variations in music, while being natural, is not often addressed as such.
As evoked previously, many generative models for music are focused on modeling the distribution of sequences and provide only ways to generate a continuation given a primer sequence \cite{sturm2016music,huang2018music,donahue2019lakhnes,Payne2019}, which is used as a conditioning. Some works focused on the so-called inpainting models for music \cite{hadjeres2018anticipation,pati2019learning} which are models able to fill in the gaps in a sequence where some tokens have been masked out to make interactive generation readily available \cite{bazin2019nonoto}. Other related approaches rely on adding pre-specified conditioning variables which allows a user to specify the current key \cite{sturm2016music,hadjeres2017deepbach}, a melody \cite{hadjeres2017deepbach,huang2019bach,choi2019encoding} or a melodic contour \cite{donahue2019piano,walder2016modelling}. On the other side of the spectrum, we can mention approaches relying on Variational Auto-Encoders (VAEs) \cite{kingma2013auto} which consist in encoding sequences of a given size into a continuous latent space. Variations are then obtained by modifying a latent code into another one before decoding. One can then have control on the modifications of the latent codes through various techniques such as following attribute vectors \cite{roberts2018hierarchical} or attribute directions \cite{hadjeres2017glsr}, performing interpolations \cite{midime}, or by learning how to slightly move in the latent space while enforcing predefined constraints \cite{engel2017latent}.
All of the above-mentioned approaches successfully produce musical results but, from the perspective of generating variations, we could say that they rely on some pre-defined similarity measure between musical pieces, either explicitly as in \cite{roy2017sampling}, or implicitly (e.g. a piece is a variation of another one if they have the exact same melody, are close in a VAE latent space, or are locally in the same key). In all cases, these notions of similarity can be either too strict (same melody) or too permissive (closeness in latent space can be unrelated to perception) and often require that the common characteristics of interest between two sequences can be computed (e.g. current key).

To overcome these limitations, we tackle the problem of producing novel sequences sharing perceptible similarities with an original template sequence \emph{without} relying on any annotation. This problem is thus intimately linked to the problem of learning meaningful high-level representations of sequences without supervision: by computing an appropriate representation of a template sequence, we can simply use this as a conditioning variable in an autoregressive generative model. The two sequences can then be considered as being similar to one another as they share the same high-level representation. 
Therefore, the crux of the problem is the following: firstly to find mechanisms to learn appropriate representations without supervision and secondly to control the amount and the nature of the information these representations contain.

For clarity, in the following, we make the extra assumption that the sequences we consider have a special structure: each sequence can be divided into natural consecutive subsequences or basic units. This assumption is often verified in practice (e.g. sequences of notes can be divided into sequences of bars, a document can be divided into a sequence of sentences) and not essential. We can then consider that a representation for such sequences can be obtained as the concatenation of the (localized) representation of its subsequences.

\begin{figure*}[h]
    \centering
    \includegraphics[scale=0.5]{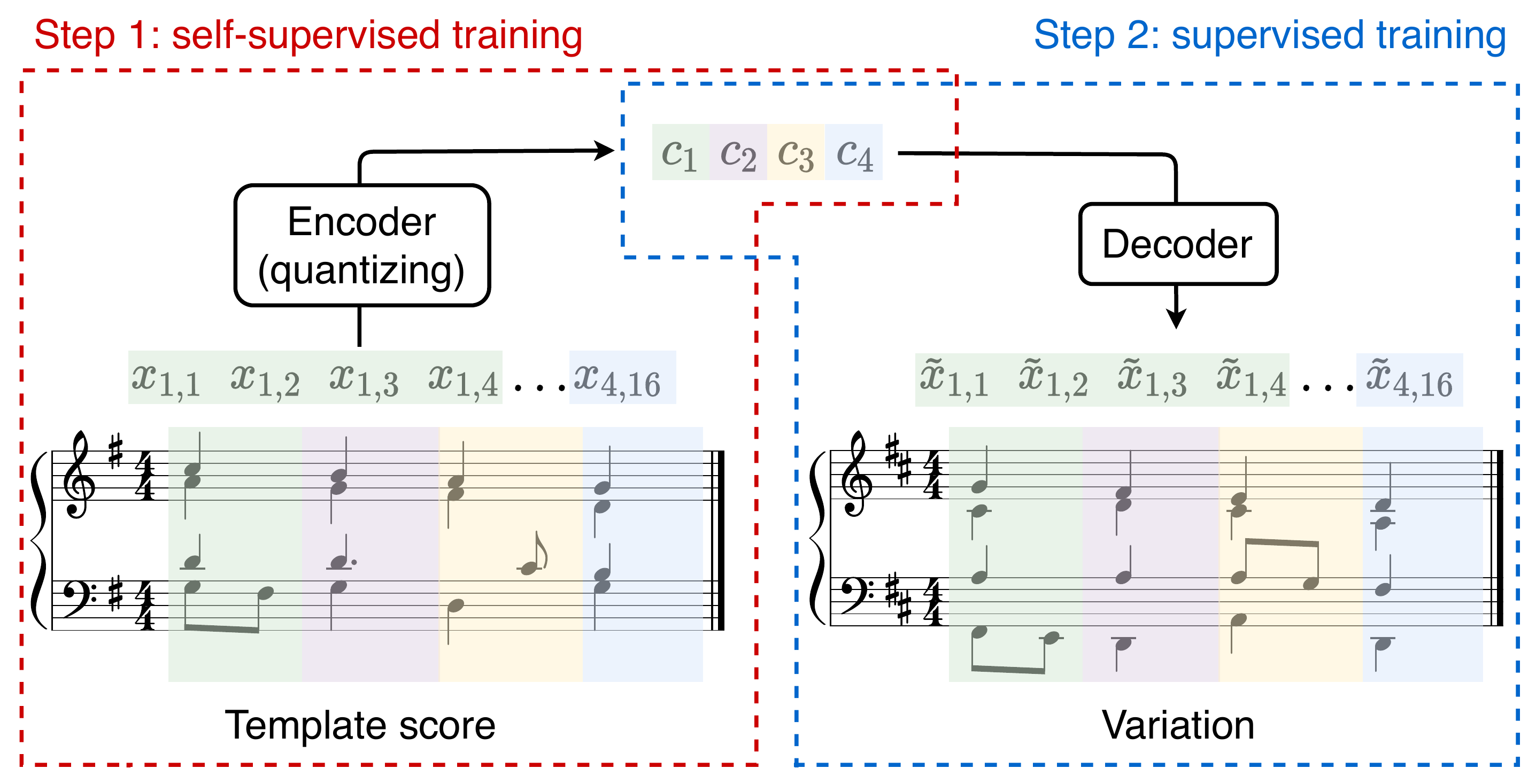}
    \caption{Schematic view of our proposed architecture comprising: An encoder trained in a self-supervised way to downsample a structured sequence $X$ into a sequence of discrete codes $c$ (red dot rectangle); a decoder trained separately in a supervised way to reconstruct a chorale $\tilde{X}$ from a sequence of codes (blue dot rectangle).
    This architecture allows us to tackle the problem of generating variations of a template sequence as a representation learning problem.
    The shaded rectangles illustrate that a latent code $c_i$ contains information strictly localised in the corresponding subsequence $x_{i}$.}
    \label{fig:global-architecture}
\end{figure*}

Our contributions are the following: 
Based on the Contrastive Predicting Coding (CPC) framework \cite{van2018cpc,henaff}, we propose \emph{Vector Quantized Contrastive Predictive Coding} (VQ-CPC), a self-supervised learning technique which consists in applying Vector Quantization \cite{van2017vqvae,de2019hierarchical} to the representations used in CPC before training them with the Noise Contrastive Estimation (NCE) loss \cite{gutmann2010noise}. Conceptually, this learns a labelling of the subsequences with discrete set of codes.
The advantages are the following: the introduction of a quantization bottleneck forces the model to map different subsequences onto the exact same codes, and thus to find common denominators and shared information among the subsequences of a same cluster;
it then becomes possible to examine the meaning of the learnt clusters (all subsequences which are mapped onto the same code). As such, VQ-CPC can be seen as a technique for unsupervised clustering and a downscaling techique mapping a sequence of discrete subsequences to a sequence of discrete codes. 
Furthermore, we show that it is possible to control the information present in the learnt clusters so that, depending on the application in mind, we can obtain different variation schemes.
We then propose to decode a sequence of codes into a sequence of subsequences using standard Transformer networks and introduce to this end an appropriate attention pattern which takes advantage of the special structure of the generated sequence.
Combined together, our proposed encoder-decoder architecture depicted in Fig.~\ref{fig:global-architecture} allows to generate variations of a given template sequence by: first extracting desired information from the template in the form of a sequence of codes, and second generate a novel sequence conditioned by that extracted sequence of codes.

We apply our method to the dataset of J.S. Bach chorales where we show that, despite the small size of the dataset, we are able to generate high-quality variations of an original sequence in different ways. Moreover, we analyse the learnt clusters from a musical point of view and demonstrate that our approach allows to capture more varied concepts than the related approach from \cite{de2019hierarchical}, which we re-implemented and adapted to symbolic music.

Even if we focused our experiments on symbolic music, our method is general and can be applied to any other data type that can be treated as a structured sequence.

The plan of this paper is the following: Section \ref{sec:background} introduces the necessary background, namely Contrastive Predictive Coding in Sect.~\ref{ssec:CPC} and Vector Quantization in Sect.~\ref{ssec:vq}. Our proposed model is introduced in Sect.~\ref{sec:model} and our experimental results on the J.S. Bach dataset are presented in Sect.~\ref{sec:experimental-results}. Lastly, Sect.~\ref{sec:related-works} compares our approach with recent related approaches. Our code is available on Github\footnote{\url{https://github.com/SonyCSLParis/vqcpc-bach}}. 
We also provide full scores and audio for all our results on the accompanying website\footnote{\url{https://sonycslparis.github.io/vqcpc-bach/}}.

{}

\section{Background}
\label{sec:background}
\subsection{Contrastive Predictive Coding}
\label{ssec:CPC}
Contrastive Predictive Coding (CPC) is a self-supervised representation learning technique relying on a contrastive objective on the latent representation itself and based on a locality criterion.
It allows to compress high-dimensional time series into a lower dimensional sequence of latent codes \cite{van2018cpc}.
Afterwards, the extracted representation can be used as an embedding of the input data on a variety of downstream tasks, and proved to reduce the training time and improve classification accuracies on classification tasks \cite{henaff,van2018cpc}.

Formally, we consider an input sequence $x = [x_1, \dots, x_L]$ split into $L$ successive non-overlapping subsequences $x_i = [x_{i1}, \dots, x_{il}]$ (possibly of different sizes).
Each subsequence $x_i$ is embedded as a real-valued vector 
\begin{equation}
\label{eq:cpc_encoder}
z_i = E_{\textrm{cpc}}(x_i) \in \mathbf{R}^d,
\end{equation}
with an encoder $E_{\textrm{cpc}}$ which is typically a recurrent neural network.
The past and present context of the embeddings $z$ at the position $i$ can be summarized into a single vector
\begin{equation}
\label{eq:cpc_h}
h_i = f_h(z_{i-K+1}, \dots , z_i) \in \mathbf{R}^d,
\end{equation}
where $K$ represents the length of the context in number of chunks and $f_h$ can be any function, typically implemented using a recurrent neural network.
This architecture is trained to maximize the mutual information between the context vector $h_i$ and the encoding of future subsequences $z_{i+k} = E_{\textrm{cpc}}(x_{i+k})$ for all $k$ in $[1, K]$.

Ideally, the objective is to maximize the mutual information between the context vector and the future latent codes, with the idea that the latent representations will embed high-level contextual information while getting rid of the local noise in the signal.
However, this quantity in intractable.
But a lower-bound can be used to derive a training criterion called the \emph{Information Noise Contrastive Estimation} (InfoNCE) loss \cite{van2018cpc}.
Using the InfoNCE loss, the model has to identify the subsequence $x_{i+k}$ (called the positive sample) effectively occurring at a later position in the sequence, from a set of $N-1$ subsequences (called negative samples) drawn from a proposal distribution, usually chosen to be uniform over the training set.
Let $\mathcal{S} = [x^{(1)}, \dots,  x^{(N)}]$ be a set containing the unique positive sample and $N-1$ negatives samples, the InfoNCE loss is defined as
\begin{equation}
\label{eq:nceloss}
\mathcal{L}_{\textrm{NCE}} = - \mathbb{E}_{S} \left[ \log \frac{f_k(x_{i+k}, h_i)}{\sum_{x_j \in \mathcal{X}} f_k(x_j, h_i)} \right],
\end{equation}
where $f_k$ is a bi-linear product
\begin{equation}
\label{eq:bilinear_product}
f_k(x_j, h_i) = \exp (z_j^T . W_k . h_i),
\end{equation}
with the $W_k$ being $k$ trainable $d \times d$  matrices.

\begin{figure*}[ht]
    \centering
    \includegraphics[scale=0.8]{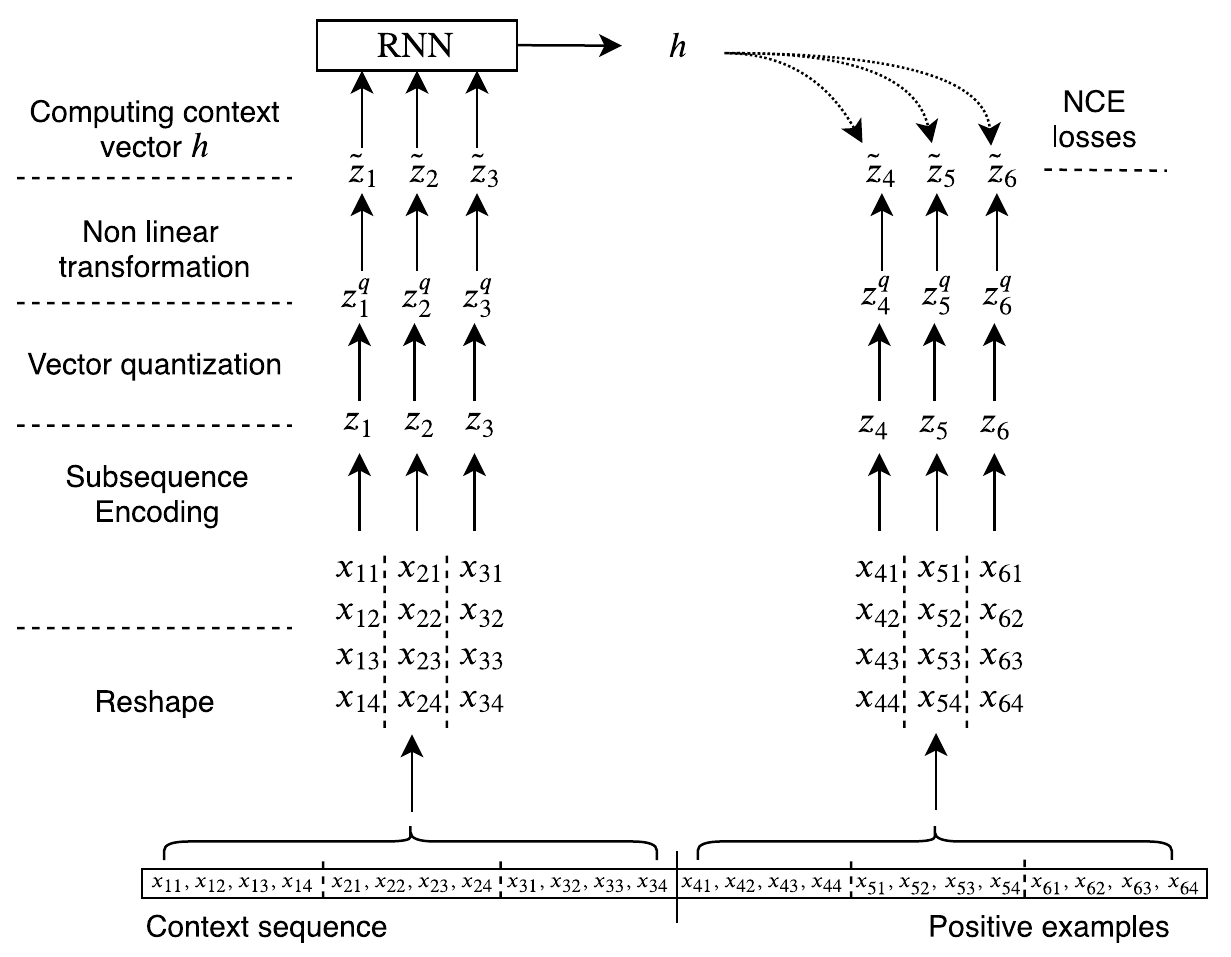}
    \caption{Schematic view of our proposed VQ-CPC training. Each subsequence is embedded in the code space before being quantized and transformed non linearly. All codes from the context sequence are then aggregated into a context vector $h$ which is used to compute the NCE losses of Eq.~\ref{eq:nceloss}.}
    \label{fig:vqcpc}
\end{figure*}{}

\subsection{Vector Quantization}
\label{ssec:vq}
Vector Quantization (VQ) \cite{Wu2019} consists in approximating the elements of a continuous vector space $\mathbb{R}^d$ by the closest element in a finite set of vectors $\mathcal{C}= \{c_1, \dots, c_C\}$ lying in the same space $\mathbb{R}^d$.
Recently, VQ has been used to quantize latent representations of auto-encoders, leading to important improvements in the quality of the generations over VAEs with a continuous latent space \cite{van2017vqvae,razavi2019generating}.

In particular, \cite{van2017vqvae} proposed a differentiable VQ layer which can easily be inserted in any model trained with gradient back-propagation.
Given a trainable set of codes $\mathcal{C}$, the quantization of an input vector $x$ is given by its closest centroid
\begin{equation}
\label{eq:quantize}
c(x) := \textrm{arg}\min_{c \in \mathcal{C}} || x - c ||_2.
\end{equation}
Because of the argmin operator, this layer is not differentiable. To circumvent that difficulty, the gradient of the loss with respect to the output of the quantization module is used as the gradient of the loss with respect to its output $\frac{\partial \mathcal{L}}{\partial x} := \frac{\partial \mathcal{L}}{\partial z^q(x)}$.
This can be implemented using the stop gradient operator~\footnote{Available in common deep learning frameworks such as PyTorch and Tensorflow}, denoted $sg$. 
Hence, for an input vector $x$, the VQ layer outputs
\begin{equation}
\label{eq:vq}
z^q(x) :=  \textrm{sg}[c(x) - x] + x,
\end{equation}

The centroid positions and the non-quantized values $z$ are updated incrementally by minimizing
\begin{equation}
\label{eq:Lvq}
\mathcal{L}_{\mathrm{VQ}} = \sum_{c \in \mathcal{C}} \delta_{z^q(x)}^c
\left(
|| \textrm{sg}[x] - c||_2 + \beta || x - \textrm{sg}[c]||_2
\right),
\end{equation}
where $\delta^a_b = 1$ iff $a=b$ and zero otherwise,  and $\beta$ 
is a parameter to control the tradeoff between the two terms. 
This loss encourages non-quantized values $x$ to be close to their assigned centroid, and is usually added to the loss function used for training the whole model.

\section{Model}
\label{sec:model}
In this section, we introduce our architecture comprised of an encoder trained in a self-supervised manner to output a down-sampled discrete representation of an input structured sequence, and a decoder trained to reconstruct a sequence from a series of discrete codes.

We consider structured sequences which consist in sequences of fixed sized subsequences of tokens.
Subsequences are typically basic structuring elements, such as beats in music or padded sentences in language.
Formally, a structured sequence defined on a finite set of symbols $\mathcal{A}$ can be represented as a matrix
$X = \left[ x_{1}, \dots, x_{L} \right] \in \mathcal{A}^{l \times L}$
where the columns $x_{i}$ represent the $L$ subsequences of identical length $l$.

An encoder $E$ separately processes each subsequence $x_i$ to assign them a discrete code $c_i = E(x_i) \in \llbracket1,C\rrbracket$.
The sequence of codes $c = [c_1, \dots, c_L]$ constitutes a lossy compression of the input sequence $X$,
and can be interpreted as an high-level representation of its structure.
A decoder $D$ is then trained in a supervised way to reconstruct the input sequence from the downsampled discrete representation $c$ extracted by the encoder.

Compared to most encoder-decoder architectures, our model is not trained in an end-to-end manner: the encoder is pre-trained independently using \emph{Vector Quantized Contrastive Predictive Coding} (VQ-CPC), a self-supervised objective which we introduce in Sect.~\ref{ssec:encoder}.
In Sect.~\ref{ssec:info-control} we explain how the information content of clusters can be controlled and in Sect.~\ref{ssec:vqcpc-decoding} we present our modelling choices for the decoder.

\subsection{Vector Quantized Contrastive Predictive Coding\label{ssec:encoder}}
VQ-CPC draws inspiration from the self-supervised learning algorithm \textit{Contrastive Predictive Coding} CPC (see Sect.~\ref{ssec:CPC}). However, the requirement that our encoder be discrete makes the direct application of this method impossible.
We propose to introduce a quantization bottleneck (see Eq.~\ref{eq:vq}) on top of the CPC encoder extracting the latent representation (see Eq.~\ref{eq:cpc_encoder}) to  obtain a discrete  latent representation of the subsequences 
\begin{equation}
E(x_i) = z^q(E_{\textrm{cpc}}(x_i)) \in \mathcal{C}
\end{equation}
where $\mathcal{C}$ is a set of $C$ centroids partitioning the embedding space $\mathcal{R}^{d_z}$.
Note that a discrete code can easily be obtained by arbitrarily indexing the centroid in $\mathcal{C}$ with an index function and using $c_i = \mathrm{index}(E(x_i))$.

The centroids purposely lie in a space with a small number of dimensions $d_z$ compared with the number of clusters $C$ to enforce a geometrical organization of the clusters.
However, this low dimensional representation is not well suited for predicting the future latent vectors $z^q$ with the simple log-bilinear product introduced in equation \ref{eq:bilinear_product}.
Hence, a non-linear transformation is applied on the quantized latent representation in order to map them to an higher dimensional space $\tilde{z}^q = \textrm{MLP}(z^q)$. We note that such a non-linear transformation of the code before the contrastive loss was also proposed in the recent \cite{chen2020simple}. This architecture is summarized in Fig.~\ref{fig:vqcpc}.

\subsection{Design choices: generating variations}
By heavily compressing the subsequences $x_i$ while maintaining good predictive performances, the encoder is forced to get rid of the local details and embed solely the higher-level contextual information.
On the other hand, a latent code $z^q_i$ only contains a very localised  information extracted from the subsequence $x_i$. 
Hence, the macro-temporal structure of the input sequence has to be preserved in the temporal organization of the sequence of latent codes.

In our setting, the quantization layer is crucial:
by limiting the information content of the codes, the decoder cannot perfectly reconstruct the original sequence, but will instead generate variations resembling the original sequence.
This aspects is demonstrated in Sect.~\ref{ssec:template_sequences}
On the other hand, a powerful enough decoder could easily achieve perfect reconstruction which would strongly hinder its ability to generate variations.
We empirically validate this intuition in Sect.~\ref{ssec:quantization_bottleneck}.

Hence, the number of clusters becomes an essential parameter as it is closely related to the amount of compression of the encoder.  We set it to a relatively small value (typically $16$ or $32$, see Sect.~\ref{sec:experimental-results}), which is one order of magnitude smaller compared with traditional VQ approaches as \cite{van2017vqvae}. Indeed, compared with these models, our objective is not to obtain crisp reconstructions but to extract meaningful high-level features.

\subsection{Controlling the information present in the latent codes}
\label{ssec:info-control}
The clustering performed by the encoder is built using the InfoNCE loss detailed in subsection \ref{ssec:CPC}.
It essentially consists in teaching a model to identify a positive example among a set of negative examples.
Hence, the features which best allow to discriminate between negative and positive examples will become the most salient features characterising the different clusters.
As a consequence, the distribution chosen to sample the negative examples will shape the information content of the clusters. 
The importance of the negative distribution as been highlighted in many works \cite{henaff,Manmatha2017}, but its impact has primarily been studied in the context of classification.
In particular, authors tend to choose a negative samples distribution that make the contrastive objective in Eq.~\ref{eq:nceloss} challenging for the encoder network.
On the contrary, our introduction of the quantization bottleneck, together with the objective of generating variations in mind, allows us to obtain interesting results using simple sampling distributions often disregarded in previous works.
We investigate different sampling schemes in Sect.~\ref{ssec:sampling_strategies} and their consequences on the learnt representations.

\subsection{Decoding VQ-CPC codes}
\label{ssec:vqcpc-decoding}

\begin{figure*}[t!]
\centering
    \includegraphics[scale=0.35]{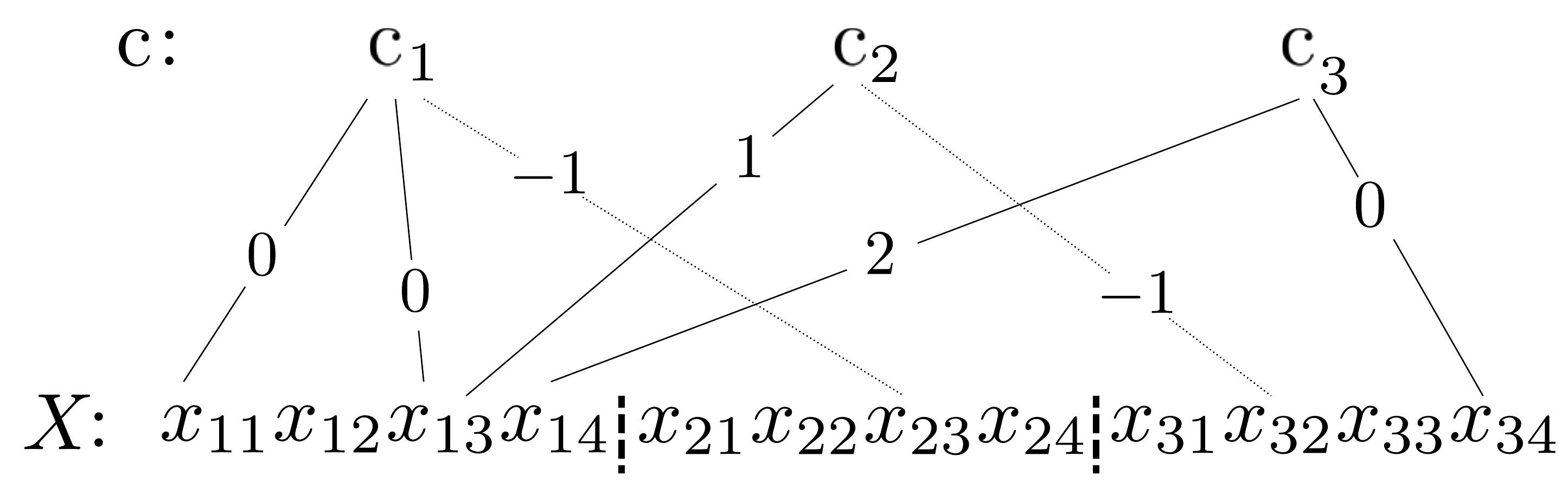}
    \caption{Example of relative distance between $X$ and a sequence of codes $c$. In our model, only the codes at a positive relative distance from a token $x_{ij}$ condition the probability of $x_{ij}$ given $x_{<ij}$.}
    \label{fig:relative-dist}
\end{figure*}

We now describe the decoder that we use to generate a sequence of subsequences $X \in \mathcal{A}^{L \times l}$ conditioned on a sequence of codes $c \in C^L$. 
For this, we use a standard seq2seq Transformer \cite{vaswani2017attention} with relative attention together with an appropriate masking scheme that takes advantage of the existing alignment between the codes $c$ and the sequences $X$.

We first note that all elements $\{x_{ij}\}_j$ of the $i^{th}$ subsequence of $X$ naturally correspond to the $i^{th}$ code $c_{i}$ of $c$ as shown in Fig.\ref{fig:relative-dist}. We use a Transformer to model the following conditional distribution:
\begin{equation}
\label{eq:conddist}
    p(X = \{x_{ij}\} | c) := \prod_{ij} p(x_{ij} | x_{<ij}, c_{\geq i}) ,
\end{equation}
where $x_{<ij} := \{x_{lm} \quad \mathrm{s.t.} \quad (l \leq i \quad  \mathrm{and} \quad m < j) \quad \mathrm{or} \quad l < i\}$ and  $c_{\geq i} = \{c_{l} \quad \mathrm{s.t.} \quad l \geq i\}$.
In particular, because of the alignment between the codes $c$ and the sequence being generated, we impose in Eq.~\ref{eq:conddist} that the element $x_{ij}$ only depends on the past elements $x_{<ij}$ and  on the \emph{present and future} codes $c_{\geq i}$ instead of the whole sequence of codes $c$.
The conditional distributions in the summation of the r.h.s. of Eq.~\ref{eq:conddist} are modeled using a transformer encoder-decoder architecture with relative attention. 
In particular, we use an anticausal mask on the encoder part (acting on codes) and a causal mask on the decoder. 
The cross-attention is a \emph{diagonal} cross-attention, meaning that when generating the token $x_{ij}$, only the output of the encoder at position $i$ is considered. 
This has the effect to speed up computation, as this does not require to perform a standard full cross-attention over all tokens, while not harming the performances of the model.

Using a causal order for generating the token $x$ but an anti-causal order for the conditioning codes $c$ is related to \cite{hadjeres2018anticipation}: it allows to prevent potential mismatch between the sequence being generated and the codes it is conditioned on when decoding. 
Furthermore, it allows to compute only once the output of the encoder part, which can then be reused during the autoregressive generation. We note that the alignment between the codes and the generated sequence allows to use relative attention on both the encoder and the decoder, which has proven to be beneficial to Transformer architectures \cite{shaw2018self}.

We use the skewing trick from \cite{huang2018music} to perform relative attention as this helps reducing the memory footprint of performing relative self-attention from $O(T^2D)$ to $O(TD)$ (where $T$ denotes the sequence length). 

\section{Experiments}
\label{sec:experimental-results}
We illustrate our approach on the dataset of J.S. Bach chorales which we present in Sect.~\ref{ssec:bach_chorales_dataset}. Implementation details about our proposed architectures are given in Sect.~\ref{ssec:implementation-details}. We compare our approach with the similar approach from \cite{de2019hierarchical} originally proposed for image generation, which can been seen as using the encoder of a VQ-VAE trained with a distilled reconstruction loss. We present how we adapted this model, which we denote as \emph{distilled VQ-VAE} in the following, to sequence generation in Sect.~\ref{sssec:distilled_vqvae}.

\subsection{Dataset}

\label{ssec:bach_chorales_dataset}
We consider the dataset of all 352 Bach chorales, which can be seen as structured sequences: based on the representation proposed in \cite{hadjeres2017deepbach},
a chorale is composed of 4 different sequences of tokens (voices)  $v^i = \{v^i_t\}_t \ , \ i \in \llbracket1, 4\rrbracket$ of equal length, where superscript $i$ indexes the voices and subscripts $t$ the time. We can then consider a flattened representation in which the four different voices are interleaved $X = [v^1_1, v^2_1, v^3_1, v^4_1, v^1_2, \dots, v^4_T]$.
By choosing a length $l$ that is a multiple of of the number of voices ($4$) for the subsequences, $X$ naturally defines a structured sequence (see Sec.~\ref{sec:model}). 
In our experiments, we considered sub-sequences of duration $1$ or $2$ beats which correspond to $16$ or $32$ tokens with a quantisation of 4 tokens per beat.
We perform data augmentations by transposing each chorale in all possible keys that respect the pitch range of the different voices.

\subsection{Implementation details}
\label{ssec:implementation-details}
\subsubsection{VQ-CPC Encoder}
\label{sssec:encoder}
The general architecture of an encoder is depicted on the left-hand side of Fig.~\ref{fig:vqcpc}.
The tokens $x_{ij}$ are first embedded in a space of dimension $32$.
Each sub-sequence $x_{i}$ is passed to a bi-directional GRU network with $2$ layers of $512$ hidden units which outputs the corresponding latent variable $z_i$.
Latent variables are quantized using $16$ clusters lying in a space of dimension $3$ to obtain $z^q_i$.
The commitment coefficient $\beta$ in Eq.~\ref{eq:Lvq} is set to $0.25$.
The clusters' centroids are initialized on a random set of $16$ examples.
A two layer MLP with hidden size $512$ is used to compute the higher dimensional $\tilde{z}^q$ whose dimension is set to $32$.
The network $f_h$ used to compute the summary vector $h$ is a $2$ layer uni-directional GRU networks with $512$ hidden layers.
The InfoNCE loss is computed on $K = 6$ successive future chunks.
A dropout value of $0.1$ is used on all layers.
The ADAM optimizer with a learning rate of $1.10^{-4}$ is used.

\subsubsection{Distilled VQ-VAE Encoder}
\label{sssec:distilled_vqvae}
The method presented in \cite{de2019hierarchical} follows the same approach as ours, in the way that an autoencoder with discrete latent variables is trained in a two-step fashion: first, an encoder which outputs discrete variables is trained on an auxiliary task and second, the decoder learns to reconstruct an input given its codes. While our auxiliary task is the minimization of the InfoNCE objective of Eq.~\ref{eq:nceloss}, this encoder is trained to minimize the following maximum likelihood-based objective: given an input sequence where a contiguous subsequence of notes has been masked, a teacher network is first trained to predict the value of the central note of this subsequence. An encoder is then obtained by training a VQ-VAE, with the difference that the reconstruction objective is replaced by a cross-entropy with the predictions given by the teacher network. We refer to this model as distilled VQ-VAE.
For our implementation, we used a bidirectional relative Transformer with $8$ layers for the teacher network and chose masks of size $8$ beats ($128$ tokens). The encoder of the VQ-VAE is a succession of 2 bidirectional relative Transformers with $4$ layers, interleaved with downscaling layers with a downscale factor of size $4$. For the downscaling layer, we simply chose a convolutional layer with a kernel size and stride of $4$. The auxiliary decoder of the VQ-VAE mirrors the structure of the encoder.
All other parameters are the same as the ones used in Sect.~\ref{sssec:encoder}.

\subsubsection{Decoder}
The decoder is a standard seq2seq Transformer \cite{vaswani2017attention} with relative self-attention as described in Sec.~\ref{ssec:vqcpc-decoding}.
The input sequences are of length $24$ beats, i.e. $384$ tokens, all embedded in a $32$-dimensional space.
Instead of directly using the centroids $z^q$, we rather use their indices in $\mathcal{C}$, and let the decoder learn a new embedding of dimensions $32$ for these indices.
Hence, the geometry of the clustering learned by the encoder is discarded when decoding and only the labels are used.
The transformer has $3$ encoder layers and $3$ decoder layers, $8$ heads of dimension $64$ and the dimension of the feedforward layers is $1028$.
As is the case in the encoder, a dropout of $0.1$ is applied to each layer and the ADAM optimizer with a learning rate of $1.10^{-4}$ is used.
Notably, training the aforementioned decoder is fast, taking approximately $1$ hour with $1$ Nvida GTX 1080 GPU.

Similarly to \cite{huang2018music}, we concatenate a positional embedding to the input of the seq2seq Transformer.
To exploit the highly structured organisation of our data we factorize our positional embeddings in two components: one accounting for the time in subdivision of the beat, and the other indexing the four voices.

\subsection{Results}
We provide scores and audio for all the results and examples mentioned in this section on the companion website of this article\footnote{\url{https://sonycslparis.github.io/vqcpc-bach/}}.

\subsubsection{Sampling strategies for shaping the clusters}
\label{ssec:sampling_strategies}
We now evaluate the impact of the sampling distribution of the negative samples on the learnt encodings as mentioned in Sect.~\ref{ssec:info-control}. An advantage when working with a small number of clusters is that we can analyze them qualitatively.
We experimented with two types of sampling for the negative samples: sampling subsequences uniformly from the dataset and sampling negative subsequences from the same (complete) sequence from which the positive sequence is drawn. In the following, we refer to these two different models as \emph{VQ-CPC-Uniform} and \emph{VQ-CPC-SameSeq} respectively.

\begin{figure*}[h!]
\centering
    \includegraphics[scale=0.096, trim=0 0 0 0, clip]{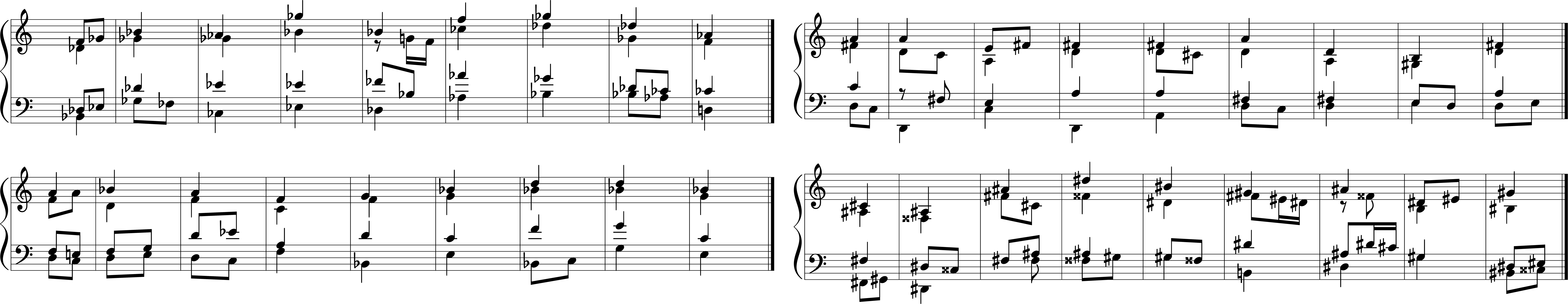}
    \caption{Four clusters obtained using the VQ-CPC-Uniform encoder. Each bar represents one subsequence and each staff one cluster.}
\label{fig:clusters-uniform}
\end{figure*}

When considering negative samples uniformly drawn from the dataset, the encoder tends to cluster subsequences according to the current key of the input. This is related to musical intuition and mimics a simple discriminative strategy that a human expert could adopt to distinguish the positive examples from the negative samples. Examples of such clusters are showed in Fig.~\ref{fig:clusters-uniform}.

\begin{figure*}[h!]
\centering
    \includegraphics[scale=0.105, trim=0 0 0 0, clip]{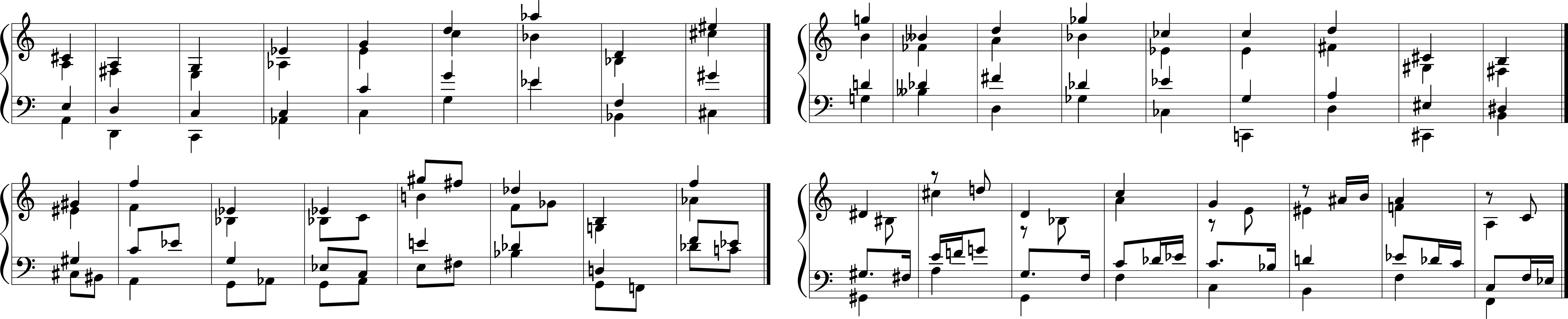}
    \caption{Four clusters obtained using the VQ-CPC-SameSeq encoder. Each bar represents one subsequence and each staff one cluster.}
\label{fig:clusters-sameseq}
\end{figure*}

When negative samples are drawn uniformly from the subsequences belonging to the same sequence from which the positive sample is drawn, the key of the input is no longer a discriminative features. This task is more difficult since using this sampling scheme resorts to teaching the network how to order a sequence.
We observe experimentally that the clusters are related to musical functions (such as cadential, pre-cadential or stable chords) and rhythmic patterns. More surprisingly, these clusters present an interesting invariance under transposition as can be seen in Fig.~\ref{fig:clusters-sameseq}.

\begin{figure*}[h!]
\centering
    \includegraphics[scale=0.5, trim=0 0 0 0, clip]{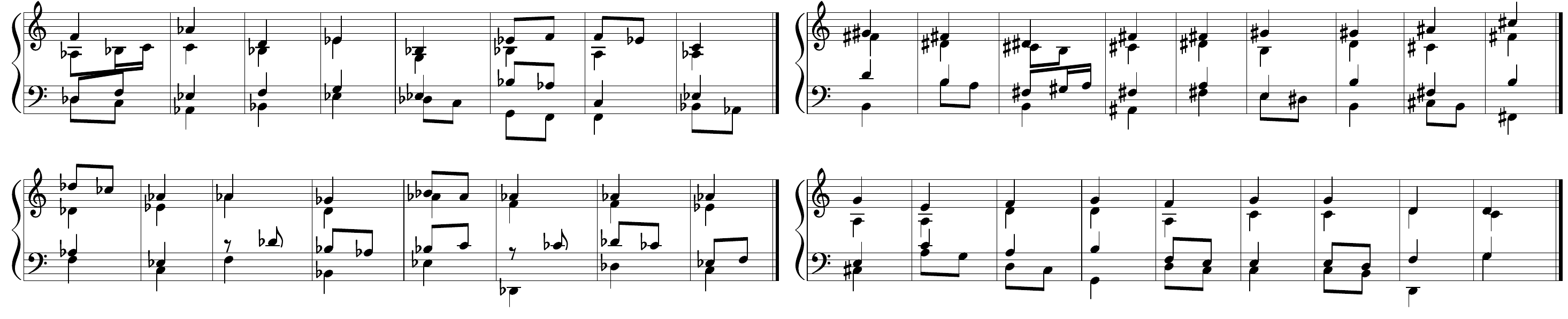}
    \caption{Four clusters obtained using the Distilled VQ-VAE encoder. Each bar represents one subsequence and each staff one cluster.}
\label{fig:clusters-distilled}
\end{figure*}

We can compare the two clustering examples above with the clustering obtained by the encoder of a Distilled VQ-VAE described in Sect.~\ref{sssec:distilled_vqvae}. By looking at Fig.~\ref{fig:clusters-distilled}, we note that the Distilled VQ-VAE encoder tends to cluster subsequences depending on their key but also depending on the soprano range. This is easily explained by the likelihood-based reconstruction objective of the Distilled VQ-VAE model. Indeed, as the teacher network is trained to predict a note given its distant neighbouring notes, its prediction are mainly based on the key of the context notes. Since the soprano line is often composed of conjunct notes in the Bach chorales dataset, it is also possible to estimate roughly the set of plausible notes for the soprano voice. These musical notions are then captured by the Distilled VQ-VAE codes.

\subsubsection{Generating variations of a template sequence}
\label{ssec:template_sequences}
Variations of sequences from the test dataset can be generated with a forward pass in the encoder-decoder as mentioned in Sect.~\ref{sec:model}. 
We used top-p sampling \cite{holtzman2019curious} with a value of 0.8 and a temperature of 0.95.

\begin{figure}[h!]
\centering
    \includegraphics[scale=0.7, trim=0 0 0 0, clip]{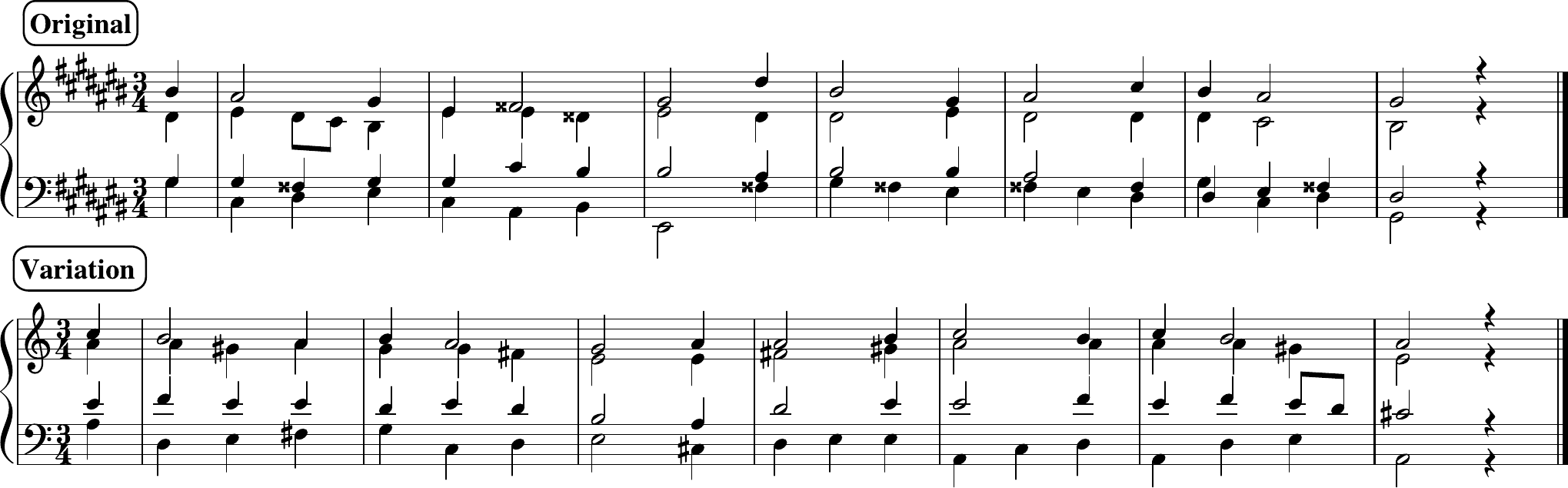}
    \caption{A variation of a template sequence  obtained by decoding VQ-CPC-SameSeq codes. The key signatures have been added manually to improve readability.}
\label{fig:variations-sameseq-main}
\end{figure}

We observe that we are able to generate new chorales that are perceptually similar from the template chorale without any plagiarism issue. The proposed variations respects the style of J.S. Bach and proved to be diverse even when using the same template sequence. In Fig.~\ref{fig:variations-sameseq-main}, we display a variation of a template sequence obtained by decoding VQ-CPC-SameSeq codes. The decoded sequence retains some meaningful musical features from the original chorale (e.g. positions of the cadences, rhythm) while discarding other features such as the key or the melodic line. It is interesting to see that this match the observations we made about the information retained in the VQ-CPC-SameSeq clusters.

In appendix, we provide more examples of variations for each of the models we considered in Sect.~\ref{ssec:sampling_strategies}, namely the VQ-CPC-Uniform (Fig.~\ref{fig:variations-random-1} and Fig.~\ref{fig:variations-random-2}), the VQ-CPC-SameSeq (Fig.~\ref{fig:variations-sameseq-1} and Fig.~\ref{fig:variations-sameseq-2}) and the Distilled VQ-VAE (Fig.~\ref{fig:variations-distilled-1} and Fig.~\ref{fig:variations-distilled-2}).

Our observation is that it is possible to obtain different variations from the same template sequence.
We also note that the importance of two different sampling strategies for the negative samples is patent and that their respective features can be clearly identified in the variations.

We also provide variations of complete chorales for the three models we considered on the accompanying website.

\subsubsection{Importance of the quantization bottleneck}
\label{ssec:quantization_bottleneck}
We observe that we are able to generate new chorales that are perceptually similar from the template chorale without any plagiarism issue.
We hypothesize that this is due to the quantization layer which limits the quantity of information contained in the codes.
To assess our hypothesis, we removed the quantization layer in our proposed architecture described in Sect.~\ref{ssec:implementation-details}.
Despite low validation error, such architecture is unable to generate diverse variations when decoding a same sequence of codes.


To assess the impact of the encoding scheme over the decoder performances, we also trained a decoder on top of an untrained VQ-CPC. Besides the higher error obtained on the test set, the model is unable to generate convincing variations, highlighting the necessity for a meaningful encoding scheme.

Due to the scarcity of available data and the relative power of the Transformer decoder we use, we believe that the quantization layer, by limiting the quantity of information embedded in the latent code, is a key component of our architecture in order to prevent over-fitting and generate diverse variations.


\section{Related Works}
\label{sec:related-works}
The idea of applying vector quantization within the CPC framework has also been proposed in the concurrent work of \cite{baevski2019vq}. In this paper, the authors use VQ-CPC to convert speech (a sequence of continuous values) into a discrete sequence. Their motivation is that discrete sequences are amenable to direct applications of Neural Language Processing algorithms. In particular, they train on top of the VQ-CPC features a BERT model \cite{devlin2018bert} in order obtain rich discrete representations for speech that proved to be useful for different classification tasks (phoneme classification and speech recognition). 
In their approach, the features extracted with VQ-CPC must retain as much information as possible about the subsequences, so that the learnt representations are applicable to a wide variety of downstream classification tasks.
Because we focus on generative applications, our requirements are different.
They report for instance the usage of potentially $102k$ different codes, which forces them to rely on standard techniques such as product quantization \cite{jegou2010product}, while we considered smaller codebook sizes of $16$ to $32$.
A notable problem when considering large codebook sizes is that there is no guarantee that all codes are used, resulting in empty clusters.
As they want to maximize the code usage, they compare different alternatives to the quantization loss of Eq.\ref{eq:Lvq}.
In our case, testing alternative quantization schemes proved to be unnecessary as our models always made use of all the codebook capacity. On the contrary, 
  they consider only one way of sampling the negative examples (drawn from the same audio example) while we were able to demonstrate and analyse the influence of the negative samples sampling scheme on the learnt representations.
  
The importance of the negative samples sampling scheme was emphasized in \cite{henaff}, where the authors demonstrated its direct influence on classifications tasks based on CPC features. An interesting aspect of this work is that they also perform data augmentation before the computation of the encodings in order to make the NCE task more challenging.
Here, we emphasize that, when introducing a quantization bottleneck, all sampling schemes can be of interest in a generative context since the learnt representations can capture different high-level aspects of the data.
In our experiments, we did not perform data augmentation. Indeed, we could have considered applying random transpositions to all subsequences before encoding them. However, we showed that, even without taking this explicitly into account, our model was still able to learn transposition-invariant clusters when negative samples were drawn from the same sequence. Nonetheless, this would be highly useful for other data modalities if one wants to explicitly disregard information during the clustering procedure. Also, an advantage of our approach is that it is possible to analyse qualitatively the significance of the learnt clusters, due to the small number of codes, without the need to train an auxiliary task based on the VQ-CPC features.

These two models have in common that they are driven by classification tasks. We showed that considering the problem of generating variations shed a different light on the same techniques. Maybe closer to our objective are the generative models based on discrete latent variables. Introduced in \cite{van2017vqvae} and improved in \cite{razavi2019generating}, these image models are auto-encoders trained with a reconstruction loss together with a quantization loss on the latent variables.
More recently, \cite{de2019hierarchical} proposed an encoder-decoder architecture where the two models are trained sequentially as in our proposed approach. The difference lies in the loss used when training the encoder. In their work, the authors propose to distill the predictions of a teacher network trained on a masked self-prediction (MSP) task to an auxiliary decoder. This has the advantage, compared to \cite{van2017vqvae,razavi2019generating}, to discard texture information in the learnt representations. Due to the similarity with our method, we re-implemented and adapted their approach to our setting (sequences of subsequences, small number of codes). This produced interesting results as the subsequences were clustered according to both the range of the soprano part and the current key. However, this method appears to be less flexible than the one proposed here since the only hyperparameter available to control the information present in the clusters is the size of the mask in the MSP task. In our experiments, we found that varying this parameter had no noticeable effect on the learnt representations. We believe that relying on a NCE loss to generate variations tends to be more appropriate than relying on losses derived from a reconstruction loss. An interesting aspect in \cite{de2019hierarchical} is that this way of constructing sequences of codes can be also applied to the sequences of codes themselves, leading to a hierarchy of discrete latent codes. Even if it is possible to apply the same idea with our model, we did not cover this aspect in this paper since the J.S. Bach dataset is quite small and generating high-quality chorales using a Transformer network is straightforward, although subject to overfitting issues.

\section{Conclusion}
We have discussed the problem of generating variations of a template sequence and casted it as a representation learning problem. We presented VQ-CPC, a self-supervised technique able to encode a long discrete sequence as a smaller sequence of interpretable discrete codes and exhibited mechanisms to control the information they carry. Finally, we proposed an attention pattern for Transformer architectures suitable for decoding efficiently a sequence of such codes and demonstrated its efficiency on the task of generating variations of J.S. Bach chorales. Our method is widely applicable and we believe that it is suitable for creative usages involving user interaction.


\bibliographystyle{abbrv}
\bibliography{vqcpc}
\appendix

\clearpage
\appendixpage
\section{Examples of variations of a template chorale}
\label{sec:appendix}
We provide more examples of variations of a template chorale obtained using the process described in Sect.~\ref{ssec:template_sequences}. 
\subsection{Decoding VQ-CPC-Uniform codes}
\begin{figure}[h!]
\centering
    \includegraphics[scale=0.5, trim=0 130 0 20, clip]{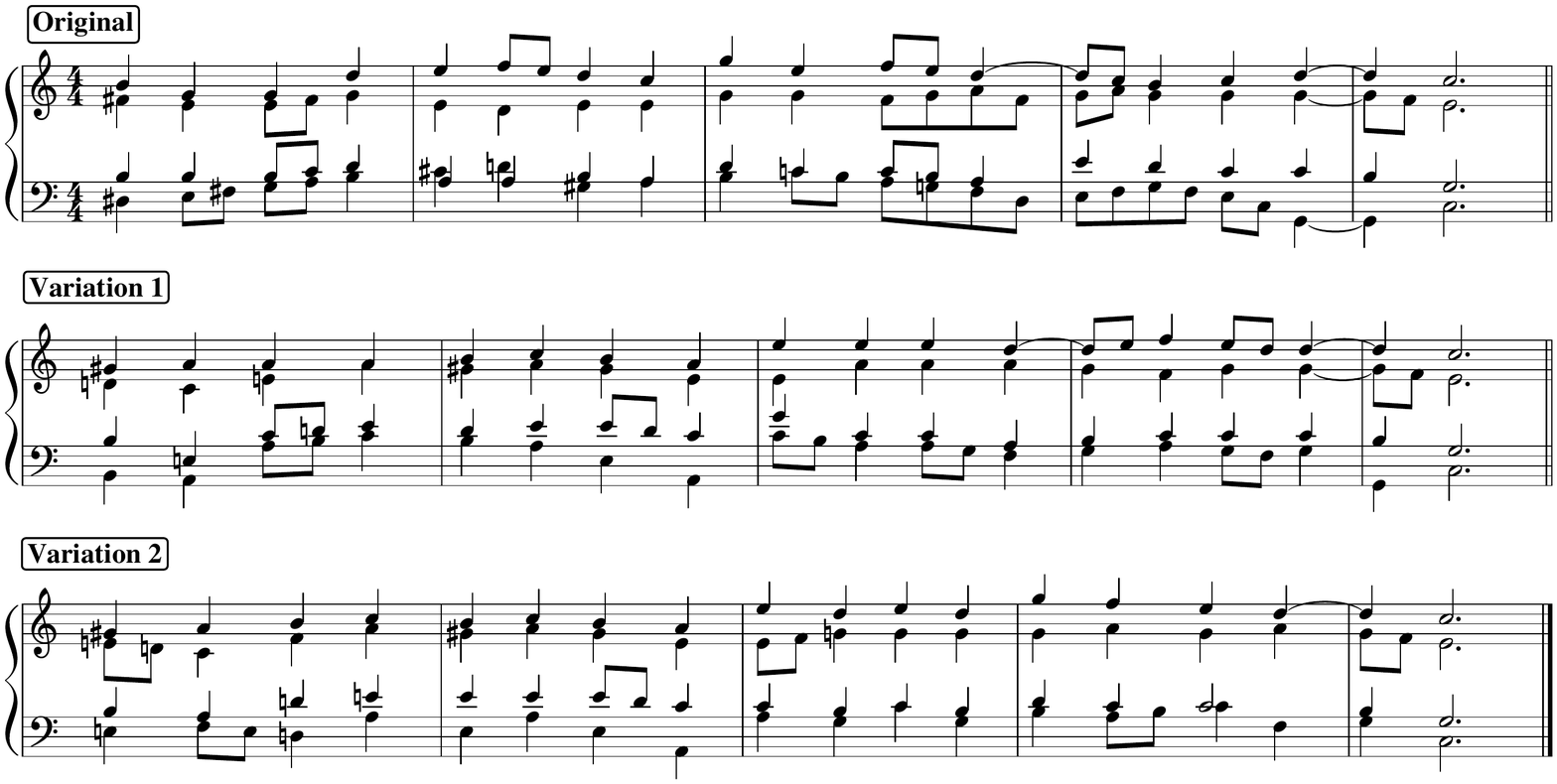}
    \caption{Two variations of a template sequence obtained by decoding its VQ-CPC-Uniform codes.}
\label{fig:variations-random-1}
\end{figure}

\begin{figure}[h!]
\centering
    \includegraphics[scale=0.5, trim=0 130 0 20, clip]{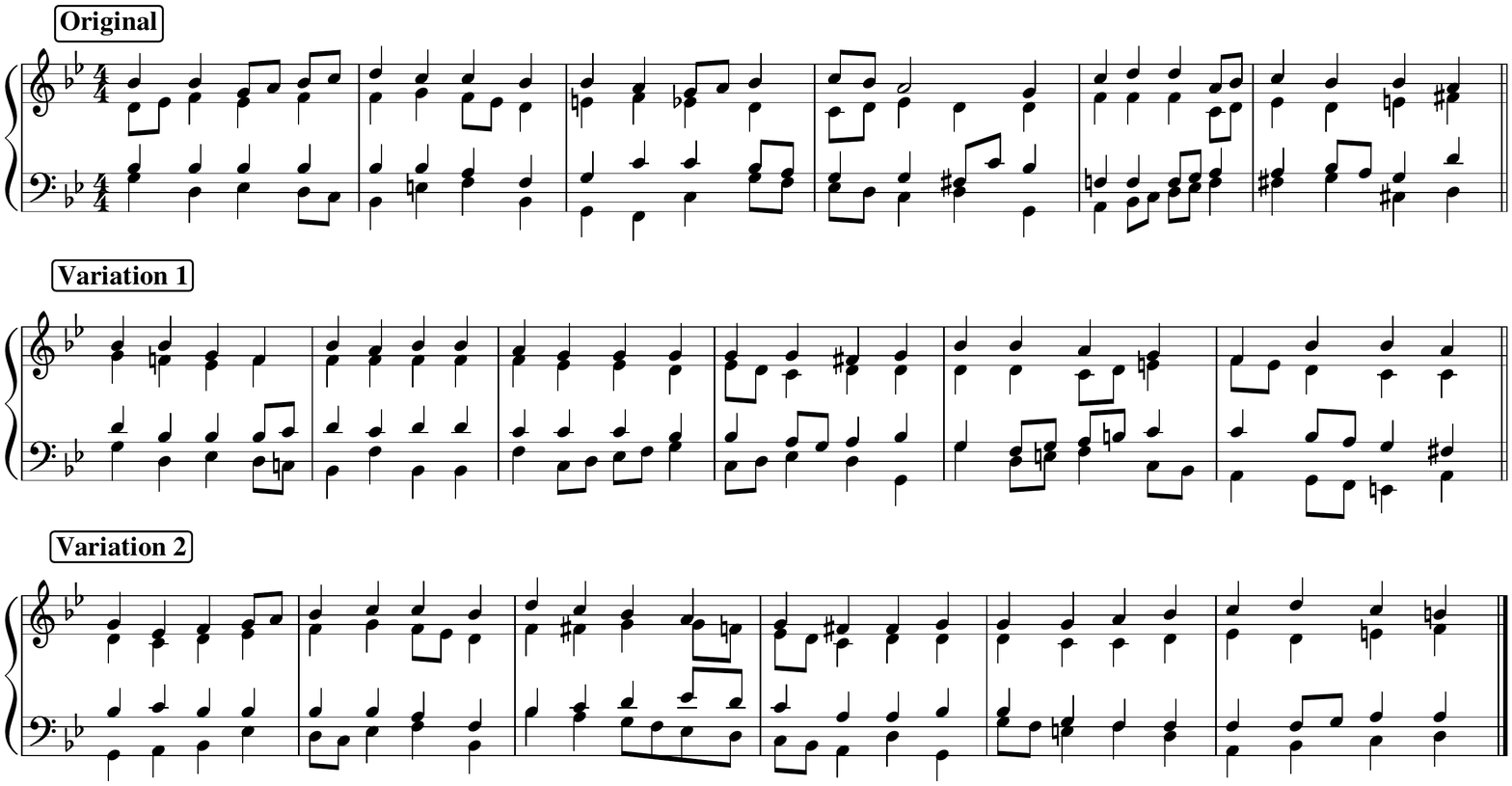}
    \caption{Two variations of a template sequence obtained by decoding its VQ-CPC-Uniform codes.}
\label{fig:variations-random-2}
\end{figure}

\clearpage
\subsection{Decoding VQ-CPC-SameSeq codes}
\begin{figure}[h!]
\centering
    \includegraphics[scale=0.5, trim=0 130 0 20, clip]{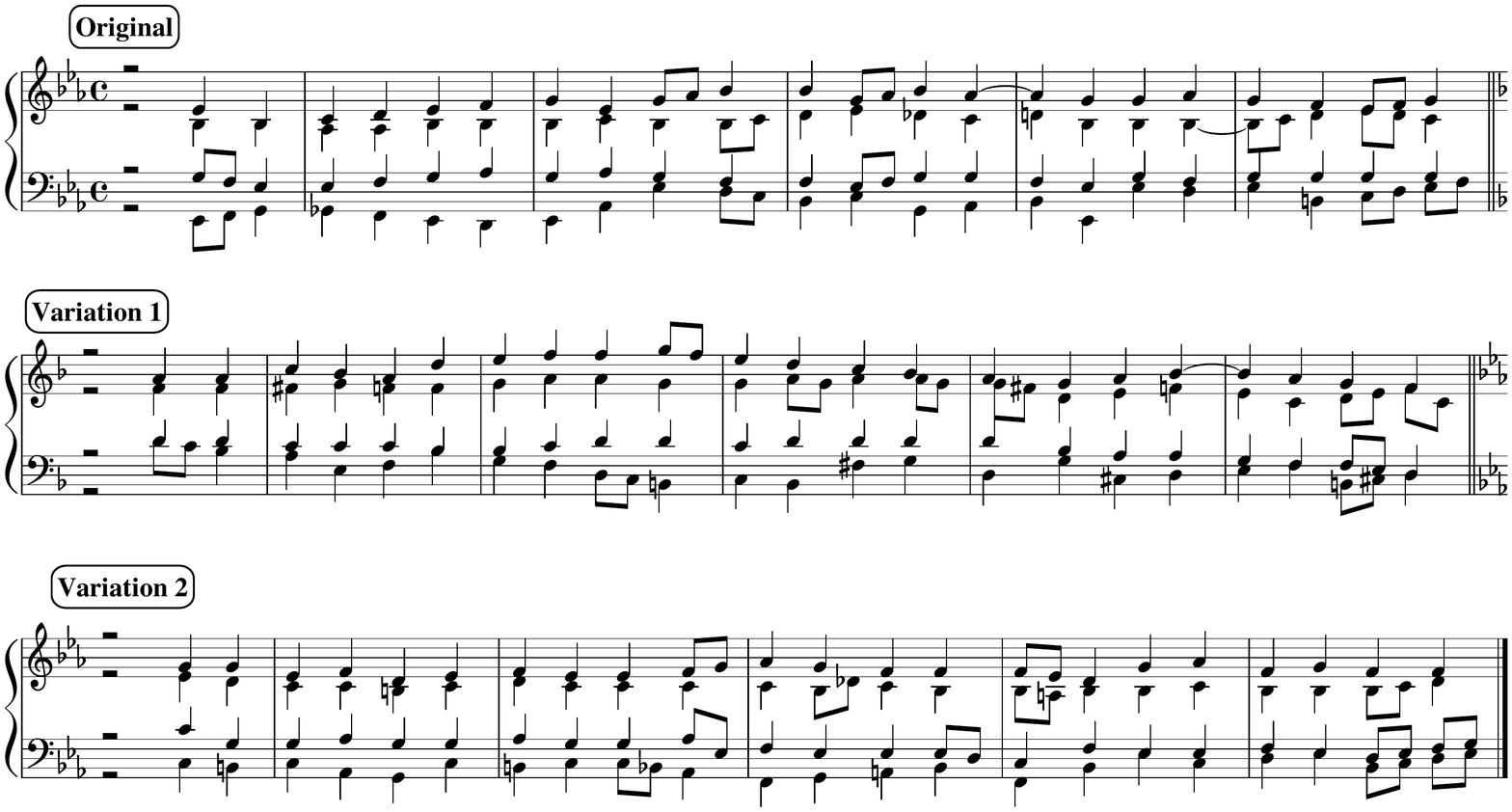}
    \caption{Two variations of a template sequence obtained by decoding its VQ-CPC-SameSeq codes.}
\label{fig:variations-sameseq-1}
\end{figure}

\begin{figure}[h!]
\centering
    \includegraphics[scale=0.5, trim=0 130 0 20, clip]{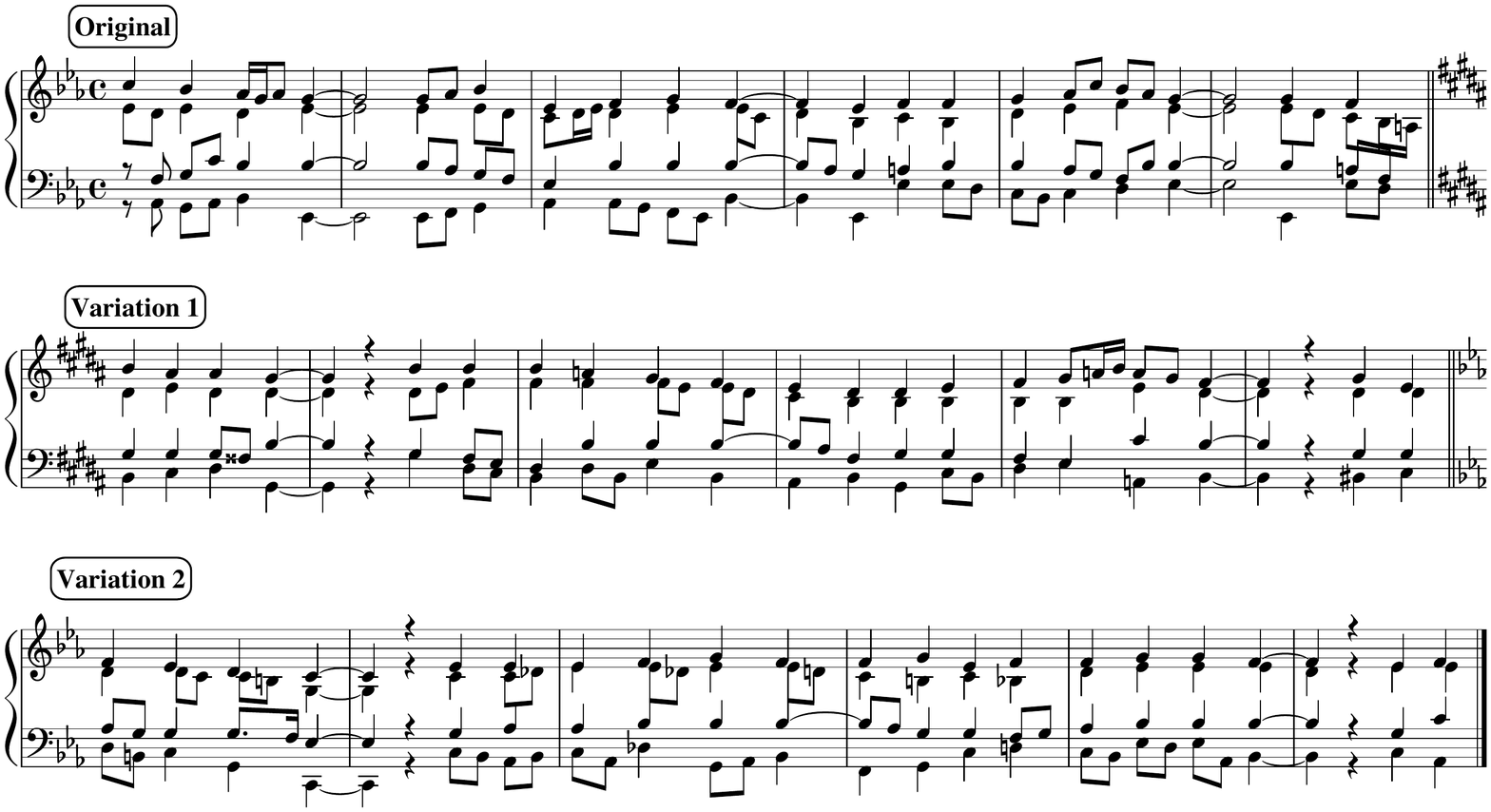}
    \caption{Two variations of a template sequence obtained by decoding its VQ-CPC-SameSeq codes.}
\label{fig:variations-sameseq-2}
\end{figure}

\FloatBarrier
\clearpage
\subsection{Decoding Distilled VQ-VAE codes}
\begin{figure}[h!]
\centering
    \includegraphics[scale=0.5, trim=0 130 0 20, clip]{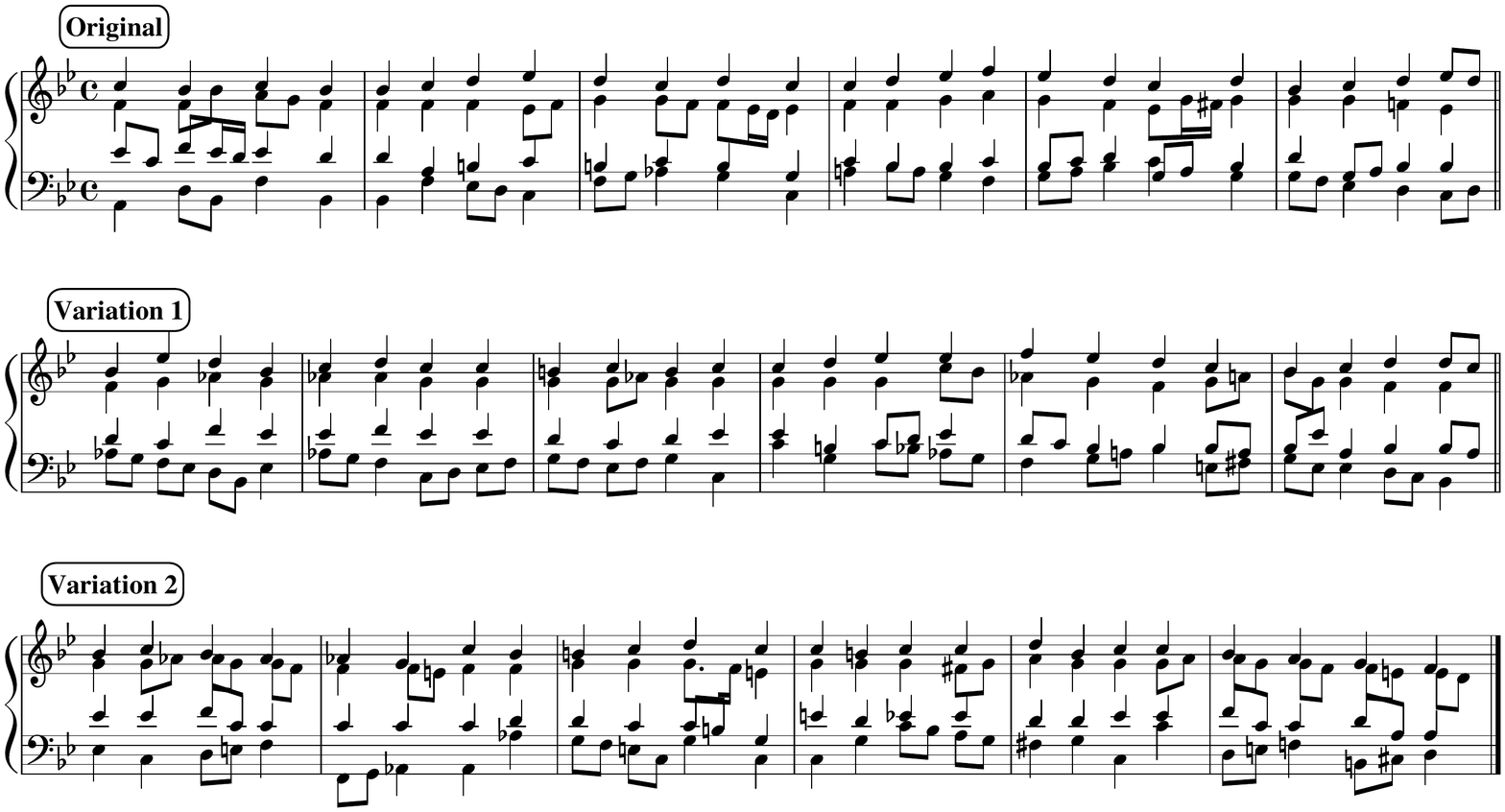}
    \caption{Two variations of a template sequence obtained by decoding its Distilled VQ-VAE codes.}
\label{fig:variations-distilled-1}
\end{figure}

\begin{figure}[h!]
\centering
    \includegraphics[scale=0.5, trim=0 130 0 20, clip]{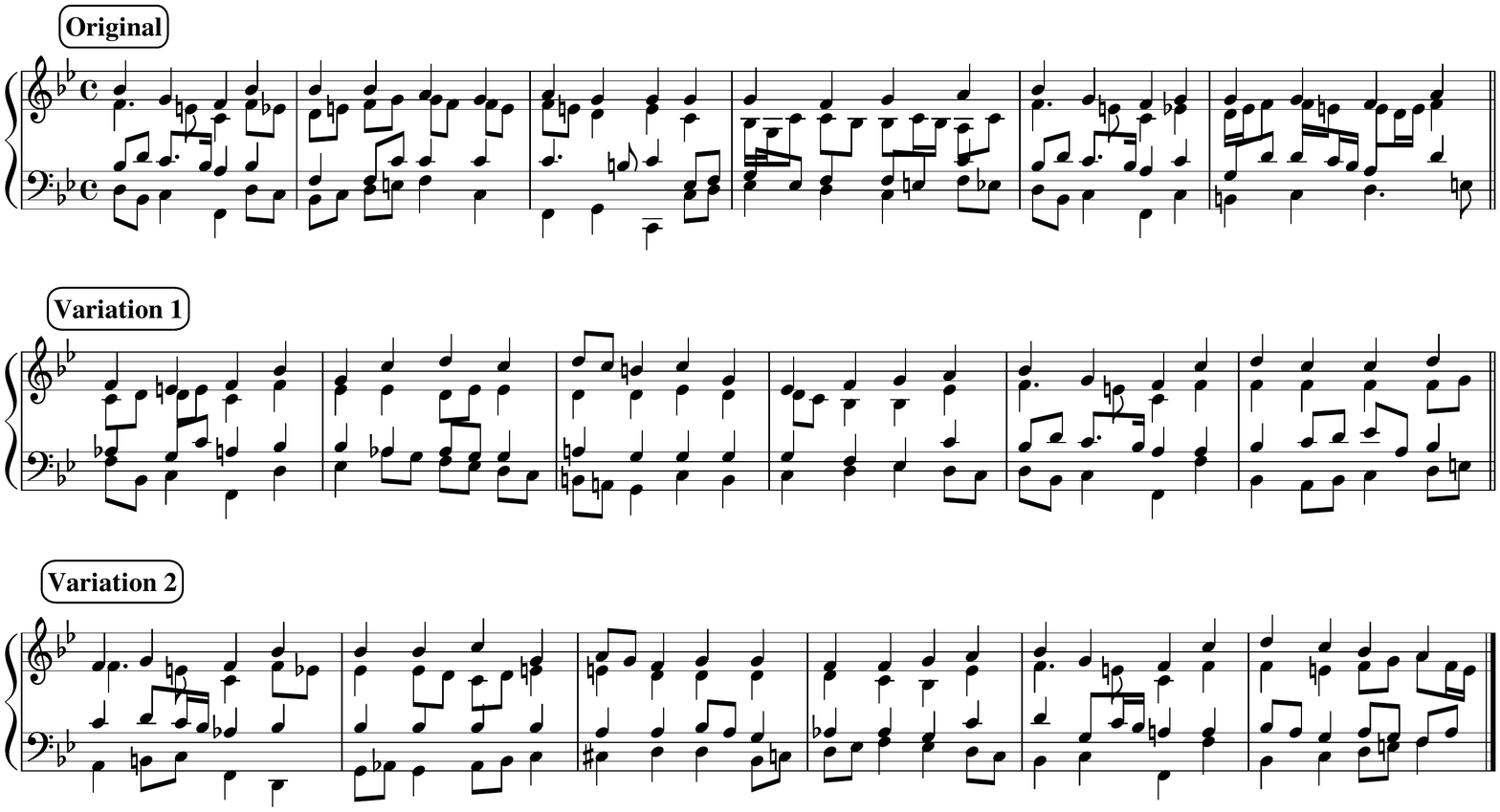}
    \caption{Two variations of a template sequence obtained by decoding its Distilled VQ-VAE codes.}
\label{fig:variations-distilled-2}
\end{figure}

\FloatBarrier
\clearpage

\end{document}